# Stack of correlated insulating states in bilayer graphene kagome superlattice

Xinyu Cai[1,2,#], Fengfan Ren[1,2,#], Qiao Li[2,3,#,*], Yanran Shi[1,2], Yifan Wang[1,2], Yifan Zhang[1,5], Zhenghang Zhi[1,5], Jiawei Luo[1,2], Yulin Chen[1,2,4], Jianpeng Liu[1,2,*], Xufeng Kou[1,2,5,*], Zhongkai Liu[1,2,*]

[1]*State Key Laboratory of Quantum Functional Materials, School of Physical Science and Technology, ShanghaiTech University, Shanghai 201210, China*

[2]*ShanghaiTech Laboratory for Topological Physics, ShanghaiTech University, 201210 Shanghai, China*

[3]*Department of Physics and Astronomy, University of California, Riverside, CA, USA.*

[4]*Department of Physics, University of Oxford, Oxford OX1 3PU, UK*

[5]*School of Information Science and Technology, ShanghaiTech University, Shanghai 201210, China*

**Graphene-based systems have emerged as a rich platform for exploring emergent quantum phenomena—including superconductivity, magnetism, and correlated insulating behavior—arising from flat electronic bands that enhance many-body interactions. Realizing such flat bands has thus far relied primarily on moiré graphene superlattices or rhombohedral stacking graphene systems, both of which face challenges in reproducibility and tunability. Here, we introduce an artificial Kagome superlattice in bilayer graphene, engineered via nanopatterning of the dielectric substrate to create a precisely defined and electrostatically tunable periodic potential. Magnetotransport measurements reveal the emergence of a *stack of correlated insulating states* at moderate superlattice potentials, characteristic of strong**

**electron–electron interactions within Kagome-induced flat bands. As temperature increases, these correlated gaps collapse, signaling the thermal suppression of interaction-driven states. Continuum-model calculations confirm the formation of multiple flat minibands and reproduce the observed evolution of band reconstruction. Our results establish dielectric-patterned graphene superlattices as a robust and controllable architecture for realizing flat-band–induced correlated phenomena beyond moiré systems.**

## Introduction

In graphene-based systems, the formation of flat electronic bands quenches kinetic energy and enhances electron–electron interactions, giving rise to symmetry-breaking ground states and exotic transport responses, including correlated insulating states, magnetism, and unconventional superconductivity[1–12]. Flat bands have been realized through two main strategies. The first employs moiré superlattices, formed by introducing small twist angles between adjacent layers or by aligning graphene with hexagonal boron nitride, yielding a periodic potential that produces flat minibands[2–4,8,9,13]. These systems, exemplified by magic-angle twisted bilayer graphene[2–4,8,9], have revolutionized flat-band research but suffer from inherent twist-angle inhomogeneity and structural relaxation. The second strategy utilizes rhombohedral stacking, where chiral interlayer coupling in ABC-stacked multilayer graphene produces flat bands[14–16]. However, this method offers only limited tunability of the band dispersion and lacks the structural design flexibility needed for exploring diverse correlated phases.

An appealing alternative is to artificially engineer superlattices that mimic the role of moiré potentials while preserving tunability (superlattice period, lattice symmetry, and crystal orientation), high quality interface and scalability[17–26], thereby allowing fabrication of high-quality devices well suited to uncovering exotic correlated phenomena. This strategy follows the general principle that periodic modulation—either arising from the substrate or introduced through patterning—can influence electronic behavior in graphene-based systems[27]. Building on this concept, dielectric-patterned graphene superlattices have been explored in several geometries[20,21,25], most notably triangular and square lattices with monolayer graphene. However, superlattices with nontrivial sublattice structures—such as the Kagome lattice—are particularly intriguing because their inherent geometrical frustration can host flat band and a wide range of correlated and topological phases. These include unconventional superconductivity[28], quantum magnetism[29], and topological edge states[30]. While preliminary results with artificial Kagome superlattices have demonstrated Dirac fermion engineering in monolayer graphene[22], clear evidence of strong electron–electron interactions and correlated quantum phases has remained elusive. To overcome these limitations, we focus on bilayer graphene (BLG), whose parabolic low-energy bands allow a Kagome superlattice to more efficiently quench kinetic energy, producing flatter and more isolated minibands. Additionally, a perpendicular displacement field can open and tune a band gap, further enhancing interaction-driven correlated states.

Here, we report the realization of a bilayer graphene (BLG) Kagome superlattice constructed via nanopatterning of a dielectric substrate, which enables high reproducibility, tunable lattice symmetry, and minimal disorder. By systematically controlling the superlattice potential, we observe band-structure modulation, most importantly, the emergence of a stack of correlated insulating states associated with Kagome-induced flat bands. These correlated states vanish at elevated temperatures as the interaction-driven gap closes. Continuum-model calculations capture the formation of multiple flat minibands and reproduce the experimentally observed evolution of the band structure. Our results establish the BLG Kagome superlattice as a robust and tunable architecture for exploring correlated phases in a controllable setting, opening new directions beyond moiré-based systems.

**Fabrication of bilayer graphene Kagome superlattice devices**

We fabricated BLG Kagome superlattice devices for transport measurements, as illustrated by the schematic and optical micrograph in Fig. 1a, b. The Kagome pattern was defined on the $SiO_2$ substrate using high-precision electron-beam lithography followed by reactive ion etching (see Methods and Supplementary Information for details). The resulting array of dielectric holes exhibits an average nearest-neighbor spacing of 52 nm, corresponding to a Kagome lattice constant of 104 nm. To preserve device quality, a thin (<5 nm) hexagonal boron nitride (hBN) flake was inserted between the patterned dielectric and the BLG[31]. Dual gating was implemented using a graphite top gate ($V_{tg}$) and Si back gate ($V_{bg}$), enabling independent control of carrier density ($n$) and the superlattice potential. Atomic force microscopy (AFM) characterization (Fig. 1c) confirms the excellent uniformity and periodicity of the patterned Kagome lattice, with uniformly distributed holes and clean surfaces. This high structural homogeneity ensures a well-defined superlattice potential in BLG, which is crucial for resolving electron–electron interaction–driven quantum phases[32].

Owing to the etched-hole dielectric structure, a global back-gate voltage $V_{bg}$ applied through the Si substrate can induce a periodic superlattice potential in BLG. As illustrated in the electrostatic model of Fig. 1d, the etched hole array modulates the local carrier density and produces a periodic variation in the on-site potential. Numerical simulations of the potential profile at $V_{bg} = -20$ V (Fig. 1e) reveal that the electrostatic potential inside the etched holes is approximately -0.3 V lower than that of the surrounding regions, thereby establishing the characteristic Kagome superlattice potential ($V_{SL}$).

**Band structure tuning in bilayer graphene Kagome superlattice**

To probe the modulation of the BLG band structure by the Kagome superlattice potential, we performed low-temperature transport measurements, with the results summarized in Fig. 2. Fig. 2a shows the longitudinal resistance ($R_{xx}$) as a function of carrier density ($n$) and $V_{SL}$. At a low $V_{SL}$ value, $R_{xx}$ exhibits a single peak at the charge neutral point, consistent with the behavior of pristine BLG. As $V_{SL}$ is increased in the negative direction, additional resistance maxima appear, and their number grows to four at $V_{SL} = -50$ V, as highlighted by the white dashed lines in Fig. 2a. In addition to these dominant features, several weaker $R_{xx}$ maxima emerge between the main peaks, signaling the development of a more complex band structure. Similar trends are also observed when $V_{SL}$ is tuned in the positive direction.

To gain further insight into the band structure modulation, we examined Landau fan diagrams at several representative values of $V_{SL}$, as indicated by the red dashed lines in Fig. 2a. At small $V_{SL}$ ($V_{SL}$ $I= 7$ V), only a single set of Landau levels emerges from the charge neutral point, with the zeroth Landau level (0th LL) marked by the green dashed line in Fig. 2b. The corresponding Hall resistance ($R_{xy}$, Fig. 2e) confirms a carrier-type transition at this point. By combining the slopes of the $R_{xx}$ minima with the associated $R_{xy}$ values (see Supplementary Information), we identify Landau level filling factors of $\pm 4$, $\pm 8$, and $\pm 12$, consistent with the characteristics of intrinsic BLG[33,34].

As $V_{SL}$ is further increased, additional $R_{xx}$ maxima appear even at relatively low magnetic fields (Fig. 2f, g), giving rise to multiple sets of Landau levels and Hofstadter butterfly patterns at higher fields, with their 0th LLs indicated by black dashed lines in Fig. 2c, d. These features bear strong resemblance to those observed in triangular and rectangular graphene superlattices[21,23,24]. However, the carrier densities associated with the $R_{xx}$ maxima at low magnetic field display unconventional behavior. Specifically, the $R_{xx}$ maxima appear at carrier densities of ~-3.0 × 10$^{11}$ cm$^{-2}$, which far exceed the expected superlattice filling density of ~4.2 × 10$^{10}$ cm$^{-2}$, calculated as $n_S = \frac{4}{A_S}$ (where 4 corresponds to four electrons per unit cell and $A_S = \frac{\sqrt{3}l_s^2}{2}$, with $l_s = 104$ nm being the Kagome lattice constant). Moreover, linecuts in Fig. 2f, g show that $R_{xy}$ does not undergo a sign reversal at these additional $R_{xx}$ maxima, except at the charge neutral point, thereafter underscoring the distinct electronic reconstruction in the Kagome geometry.

The unconventional transport behavior observed in the BLG Kagome superlattice highlights the emergence of a unique band structure, distinct from that in triangular or square graphene superlattices[21,23,24]. First, the appearance of minor $R_{xx}$ maxima between the main resistance peaks indicates the presence of additional subtle band features beyond simple band gaps, consistent with the formation of a series of density-of-states (DOS) minima. Unlike triangular or square superlattices[21,23,24], no well-resolved Landau levels emerge from these secondary $R_{xx}$ maxima, implying strong hybridization between adjacent bands that blurs the gap edges and replaces sharp band gaps with shallow DOS minima.

Second, the carrier densities corresponding to the high-$V_{SL}$ $R_{xx}$ maxima are significantly larger than the expected superlattice filling density ($n_S$). This stands in sharp contrast to triangular or rectangular superlattices, where higher-order Dirac points appear, yielding $R_{xx}$ maxima at integer multiples of $n_S$. We attribute this deviation to the more intricate electronic reconstruction in the Kagome geometry, where neither a single reconstructed Dirac point nor a well-isolated band corresponding to integer multiples of $n_S$ is formed.

Third, the Hall response near the additional $R_{xx}$ maxima exhibits anomalies. Unlike at the charge neutral point or reconstructed Dirac points, $R_{xy}$ does not undergo a sign reversal but instead shows a steepened slope, indicating a substantial redistribution of carriers among multiple Fermi pockets. This behavior reflects the complex multi-component nature of the Fermi surface arising from the strongly reconstructed Kagome-modulated bandstructure.

**Observation of correlated insulating states in bilayer graphene Kagome superlattice**

In addition to the pronounced transport anomalies observed at large superlattice potentials, we examined the regime of moderate $V_{SL}$, where clear signatures of strong electron–electron correlations emerge. Figure 3a presents a high-resolution $R_{xx}$ map extracted from Fig. 2b, focusing on a narrow carrier-density range around $-7 \times 10^{11}$ cm$^{-2}$. The corresponding differential map of $R_{xx}$ along the carrier-density axis (Fig. 3b), obtained after background subtraction via polynomial fitting, reveals a series of multiple parallel $R_{xx}$ minima. Horizontal linecuts taken at successive magnetic fields (Fig. 3c) show that these minima follow nearly linear trajectories with identical slopes as the magnetic field increases, as highlighted by the dashed guide lines. The carrier densities corresponding to these $R_{xx}$ minima exhibit pronounced periodicity, manifested as uniform spacing between adjacent minima of Fig. 3c.

To quantify this periodicity, we performed a fast Fourier transform (FFT) of the carrier-density–dependent oscillations from Fig. 3b. The resulting FFT spectrum (Fig. 3d) displays four prominent peaks, marked in different colors, corresponding to carrier-density periods of $2.2 \times 10^{10}$, $3.2 \times 10^{10}$, $4.1 \times 10^{10}$, and $6.1 \times 10^{10}$ cm$^{-2}$. These periodicities correspond respectively to $\frac{n_s}{2}$, $\frac{3n_s}{4}$, $n_s$, and $\frac{3n_s}{2}$, where $n_s$ denotes the carrier density required to fill the Kagome superlattice with four electrons per unit cell.

The observed carrier-density periodicity at integer multiples of $\frac{n_s}{4}$ indicates that the series of parallel $R_{xx}$ minima originates from correlated electron–electron interactions, which we interpret as signatures of a **stack of correlated insulating states**. By tracing the $R_{xx}$ minima in Fig. 3b (highlighted by dashed lines in Fig. 3e) and analyzing their slopes, we extract the corresponding Landau-level filling factors using the relation $\Delta n/n_0 = \nu \Delta\phi/\phi_0$ (Fig. 3f). The histogram of extracted values (inset, Fig. 3f) yields a mean filling factor of $\nu = -4$, confirming that all $R_{xx}$ minima correspond to Landau levels with the same filling factor. The uniform slope of –4 across the series further indicates that these states share identical filling factor but occur at distinct carrier densities.

The periodic spacing of the $R_{xx}$ minima at integer multiples of $\frac{n_s}{4}$ implies that the corresponding Landau levels originate from a sequence of carrier densities separated by well-defined intervals. Extrapolating these traces to zero magnetic field suggests the presence of multiple $R_{xx}$ maxima corresponding to a hierarchy of insulating states. The insulating states with a periodicity of $n_s$ can be attributed to superlattice-induced band gaps, whereas those at fractional multiples of $n_s$ are likely driven by symmetry-breaking effects arising from strong electron–electron interactions[3,7–9].

Furthermore, additional Landau levels with filling factors of –8, +4, and +8 emerge from the same carrier densities as the zero-field $R_{xx}$ maxima (see Supplementary Information), supporting the coexistence of both band-gap insulating states and correlated insulating states. These features become well-resolved only at high magnetic fields and are gradually smeared out as the field decreases, consistent with the reduced Landau-level energy spacing in the low-field regime.

Transport measurements at elevated temperature further substantiate the coexistence of band-gap and correlated insulating states in the BLG Kagome superlattice. As shown in Fig. 4a–c, the series of linear $R_{xx}$ minima traces persist at 10 K but exhibit increased spacing between adjacent features compared with the 100 mK data. FFT analysis of the carrier-density oscillations (Fig. 4d) reveals that only a single periodic component remains, with a carrier-density period of $(4.3 \pm 0.2) \times 10^{10}$ cm$^{-2}$—close to the superlattice filling density $n_s$. The corresponding Landau-level filling factors remain at ν = –4. These results indicate that the correlated-induced insulating states possess smaller energy gaps than the superlattice band-gap states and are thus more easily quenched by thermal excitation, consistent with the closing of the interaction-driven gap at elevated temperatures.

Importantly, these correlated insulating states are reproducible in a second BLG device with the same Kagome-patterned superlattice, as well as in an additional device with a smaller lattice constant of 88 nm. In both cases, the observed features are consistent with those in the main device, demonstrating that the correlated insulating behavior is an intrinsic characteristic of bilayer graphene Kagome superlattices (see Supplementary Information).

**Origin of correlated insulating states**

To further elucidate the band-structure modulation in BLG under a Kagome superlattice potential and clarify the origin of the observed stack of correlated insulating states, we performed continuum-model calculations[35,36]. The total Hamiltonian of the BLG Kagome superlattice can be expressed as

$$H = H_{bi} + U_{SL}.$$

where $H_{bi}$ represents the intrinsic BLG Hamiltonian and $U_{SL}$ denotes the electrostatic superlattice potential, which is experimentally tuned via the back-gate voltage ($V_{bg}$). Variations in $V_{bg}$ modify

$U_{SL}$ and consequently reshape the electronic band structure.

As shown in Fig. 5a, at a low $V_{bg}$, the calculations reveal the emergence of a series of well-isolated flat bands (highlighted in red) near the charge neutrality point, with bandwidths below 5 meV. These flat bands originate from band folding and reconstruction induced by the Kagome superlattice potential, leading to electron–electron interactions. The enhanced interactions spontaneously break spin–valley symmetry within each flat band[3,7–9,37], giving rise to the experimentally observed stack of correlated insulating states (Fig. 3).

Upon increasing $V_{bg}$, the superlattice modulation strengthens, producing more pronounced band reconstruction (Fig. 5b, c). Although the low-energy bands remain relatively narrow, they begin to overlap, reducing the effective correlation strength. Meanwhile, larger energy gaps open at higher energies compared with the weak-modulation regime (highlighted by the red shaded regions), consistent with the emergence of additional $R_{xx}$ peaks observed at zero magnetic field in Figs. 2c and 2d. The enhanced band overlap at strong modulation also explains the progressive blurring of the Landau fan patterns at high $V_{bg}$, where Landau levels become poorly defined—consistent with the diminished $R_{xx}$ minima observed at low magnetic fields.

**Conclusions**

In summary, we have realized a BLG Kagome superlattice by nanopatterning a dielectric substrate and demonstrated gate-tunable band reconstruction controlled by the applied superlattice potential. At moderate modulation strength, transport measurements reveal the emergence of a stack of correlated insulating states originating from Kagome-induced flat bands, in excellent agreement with continuum-model calculations. These results establish the BLG Kagome superlattice as a reproducible and highly tunable platform for investigating strong electron–electron correlations in two dimensions. Beyond advancing flat-band physics, this artificial lattice approach provides a versatile and scalable route to engineer correlated, magnetic, and topological quantum phases in graphene-based materials, bridging the gap between moiré engineering and lithographically defined quantum architectures.

**Method**

**Pattern fabrication and device assembly**

Kagome superlattices were fabricated with electron beam lithography (EBL) and etching. First, we wrote the Kagome pattern of an array of circular holes with a diameter 36 nm and periodicity 104 nm on $SiO_2$/Si substrate coated with PMMA A2. After the development, the exposed $SiO_2$ regions were etched by reactive ion etching (RIE) with $CF_4$ gas to a depth of approximately 30 nm. Finally, we used Oxygen plasma to remove residual resist and contaminants. Thus, we can obtain relative clean and uniform Kagome superlattice shown in Fig. 1c.

Bilayer graphene (BLG) encapsulated with hexagonal boron nitride (hBN) was assembled onto the patterned substrate via a dry transfer technique. The top hBN thickness was maintained near 20 nm, while the bottom hBN was kept below 5 nm to enhance the modulation amplitude of the superlattice potential. We use graphite flake as the top gate, and heavily doped Si substrate as the global back gate. After the whole stacking processing, the heterostructure was patterned into a multi-terminal Hall bar geometry and then evaporated with Cr/Au as edge contacts.

**Transport measurements**

Low-temperature transport measurements were carried out in a 14 T Oxford dilution refrigerator with a base temperature of 100 mK. Resistance was measured using a lock-in amplifier (Stanford Research Systems SR830) with an excitation current of 100 nA root mean square (RMS) at a frequency of 17.777 Hz. Top-gate and back-gate voltages were applied via Keithley 2400 source meters.

**Data availability**

The data supporting the findings of this study are available from the corresponding authors upon request.

**Reference**

1. Sun, X. *et al.* Correlated states in doubly-aligned hBN/graphene/hBN heterostructures. *Nat. Commun.* **12**, 7196 (2021).

2. Cao, Y. *et al.* Tunable correlated states and spin-polarized phases in twisted bilayer–bilayer graphene. *Nature* **583**, 215–220 (2020).

3. Cao, Y. *et al.* Correlated insulator behaviour at half-filling in magic-angle graphene superlattices. *Nature* **556**, 80–84 (2018).

4. Cao, Y. *et al.* Unconventional superconductivity in magic-angle graphene superlattices. *Nature* **556**, 43–50 (2018).

5. Lu, X. *et al.* Superconductors, orbital magnets and correlated states in magic-angle bilayer graphene. *Nature* **574**, 653–657 (2019).

6. Pierce, A. T. *et al.* Unconventional sequence of correlated Chern insulators in magic-angle twisted bilayer graphene. *Nat. Phys.* **17**, 1210–1215 (2021).

7. He, M. *et al.* Symmetry breaking in twisted double bilayer graphene. *Nat. Phys.* **17**, 26–30

(2021).

8. Shen, C. *et al.* Correlated states in twisted double bilayer graphene. *Nat. Phys.* **16**, 520–525 (2020).

9. Xie, M. & MacDonald, A. H. Nature of the Correlated Insulator States in Twisted Bilayer Graphene. *Phys. Rev. Lett.* **124**, 097601 (2020).

10. Choi, Y. *et al.* Correlation-driven topological phases in magic-angle twisted bilayer graphene. *Nature* **589**, 536–541 (2021).

11. Mao, J. *et al.* Evidence of flat bands and correlated states in buckled graphene superlattices. *Nature* **584**, 215–220 (2020).

12. Saito, Y. *et al.* Hofstadter subband ferromagnetism and symmetry-broken Chern insulators in twisted bilayer graphene. *Nat. Phys.* **17**, 478–481 (2021).

13. Bistritzer, R. & MacDonald, A. H. Moiré bands in twisted double-layer graphene. *Proc. Natl. Acad. Sci.* **108**, 12233–12237 (2011).

14. Han, T. *et al.* Correlated insulator and Chern insulators in pentalayer rhombohedral-stacked graphene. *Nat. Nanotechnol.* **19**, 181–187 (2024).

15. Zhang, Y. *et al.* Layer-dependent evolution of electronic structures and correlations in rhombohedral multilayer graphene. *Nat. Nanotechnol.* **20**, 222–228 (2025).

16. Liu, K. *et al.* Spontaneous broken-symmetry insulator and metals in tetralayer rhombohedral graphene. *Nat. Nanotechnol.* **19**, 188–195 (2024).

17. Sandner, A. *et al.* Ballistic Transport in Graphene Antidot Lattices. *Nano Lett.* **15**, 8402–8406 (2015).

18. Jessen, B. S. *et al.* Lithographic band structure engineering of graphene. *Nat. Nanotechnol.* **14**,

340–346 (2019).

19. Wang, L. *et al.* Quantum transport properties of monolayer graphene with antidot lattice. *J. Appl. Phys.* **126**, 084305 (2019).

20. Li, Y. *et al.* Anisotropic band flattening in graphene with one-dimensional superlattices. *Nat. Nanotechnol.* **16**, 525–530 (2021).

21. Forsythe, C. *et al.* Band structure engineering of 2D materials using patterned dielectric superlattices. *Nat. Nanotechnol.* **13**, 566–571 (2018).

22. Wang, S. *et al.* Dispersion-Selective Band Engineering in an Artificial Kagome Superlattice. *Phys. Rev. Lett.* **133**, 066302 (2024).

23. Barcons Ruiz, D. *et al.* Engineering high quality graphene superlattices via ion milled ultra-thin etching masks. *Nat. Commun.* **13**, 6926 (2022).

24. Huber, R. *et al.* Gate-tunable two-dimensional superlattices in graphene. *Nano Lett.* **20**, 8046–8052 (2020).

25. Mreńca-Kolasińska, A., Chen, S.-C. & Liu, M.-H. Probing miniband structure and Hofstadter butterfly in gated graphene superlattices via magnetotransport. *Npj 2D Mater. Appl.* **7**, 64 (2023).

26. Huber, R. *et al.* Band conductivity oscillations in a gate-tunable graphene superlattice. *Nat. Commun.* **13**, 2856 (2022).

27. Ren, J. *et al.* Kondo Effect of Cobalt Adatoms on a Graphene Monolayer Controlled by Substrate-Induced Ripples. *Nano Lett.* **14**, 4011–4015 (2014).

28. Ortiz, B. R. *et al.* $CsV_3Sb_5$ : A $Z_2$ Topological Kagome Metal with a Superconducting Ground State. *Phys. Rev. Lett.* **125**, 247002 (2020).

29. Yin, J.-X. *et al.* Giant and anisotropic many-body spin–orbit tunability in a strongly correlated

kagome magnet. *Nature* **562**, 91–95 (2018).

30. Kang, M. *et al.* Dirac fermions and flat bands in the ideal kagome metal FeSn. *Nat. Mater.* **19**, 163–169 (2020).

31. Dean, C. R. *et al.* Boron nitride substrates for high-quality graphene electronics. *Nat. Nanotechnol.* **5**, 722–726 (2010).

32. Kammarchedu, V., Butler, D., Rashid, A. S., Ebrahimi, A. & Kayyalha, M. Understanding disorder in monolayer graphene devices with gate-defined superlattices. *Nanotechnology* **35**, 495701 (2024).

33. Zhao, Y., Cadden-Zimansky, P., Jiang, Z. & Kim, P. Symmetry Breaking in the Zero-Energy Landau Level in Bilayer Graphene. *Phys. Rev. Lett.* **104**, 066801 (2010).

34. Yagi, R. *et al.* Low-energy band structure and even-odd layer number effect in AB-stacked multilayer graphene. *Sci. Rep.* **8**, 13018 (2018).

35. Moon, P. & Koshino, M. Optical absorption in twisted bilayer graphene. *Phys. Rev. B* **87**, 205404 (2013).

36. McCann, E. & Koshino, M. The electronic properties of bilayer graphene. *Rep. Prog. Phys.* **76**, 056503 (2013).

37. Wang, Z.-Y. *et al.* Visualizing localized nematic states in twisted double bilayer graphene. *Nanoscale* **16**, 18852–18858 (2024).

## Acknowledgments

Z.K.L. acknowledges support from the National Natural Science Foundation of China (92365204, 12274298) and National Key R&D Program of China (Grant No. 2022YFA1604400/03). Q.L.

acknowledges support from the National Natural Science Foundation of China (Grant No. 12304231).

## Author contributions

Q.L. and Z.K.L. conceived of and designed the experiments. X.Y.C. prepared the BLG Kagome superlattices. X.Y.C., F.F.R and Q. Li. performed the transport measurement and data analysis with the help of X.F.K. Y.R.S, F.F.R. and J.P.L. performed theoretical analysis. X.Y.C., Q.L. and Z.K.L. wrote the manuscript with input from all authors.

## Competing interests

The authors declare no competing interests.

## Additional information

**Supplementary information** is available for this paper at the URL inserted when published.

**Correspondence** and requests for materials should be addressed to Q.L.(qiao.li1@ucr.edu), J.P.L.(liujp@shanghaitech.edu.cn), X.F.K.(kouxf@shanghaitech.edu.cn) and Z.K.L.(liuzhk@shanghaitech.edu.cn).

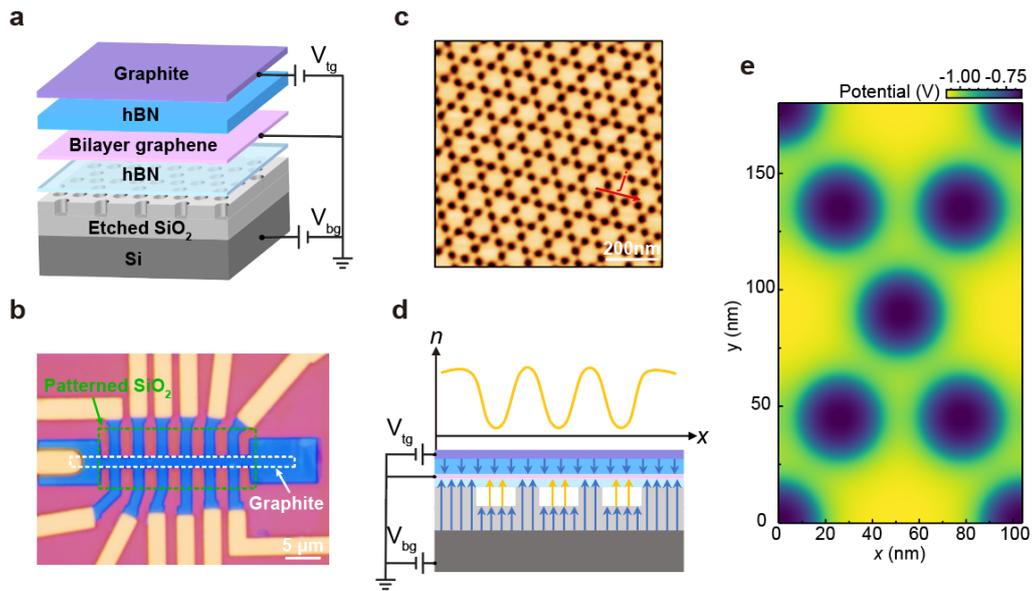

**Figure 1 | Device architecture and characterization of the bilayer graphene Kagome superlattice. a**, Schematic illustration of the bilayer graphene (BLG) Kagome superlattice device. The superlattice potential is induced by a nanopatterned SiO$_2$/Si substrate with a period of 104 nm. **b**, Optical micrograph of the fabricated BLG device in a multi-terminal Hall-bar geometry. The white dashed outline marks the graphite top-gate region, and the green dashed outline indicates the patterned SiO$_2$ area. Scale bar, 5 μm. **c**, Atomic-force microscopy image of the patterned SiO$_2$ substrate, showing a uniform Kagome-lattice array of etched holes. The red arrow denotes the current-flow direction in the transport channel. Scale bar, 200 nm. **d**, Schematic of the dual-gate electrostatic configuration. Carrier density and superlattice potential are independently tuned via the graphite top gate ($V_{tg}$) and Si back gate ($V_{bg}$). **e**, Simulation of electrostatic potential distribution at $V_{bg} = -20$ V, showing lower potential inside the etched holes relative to surrounding regions, forming the Kagome-lattice superlattice potential $V_{SL}$.

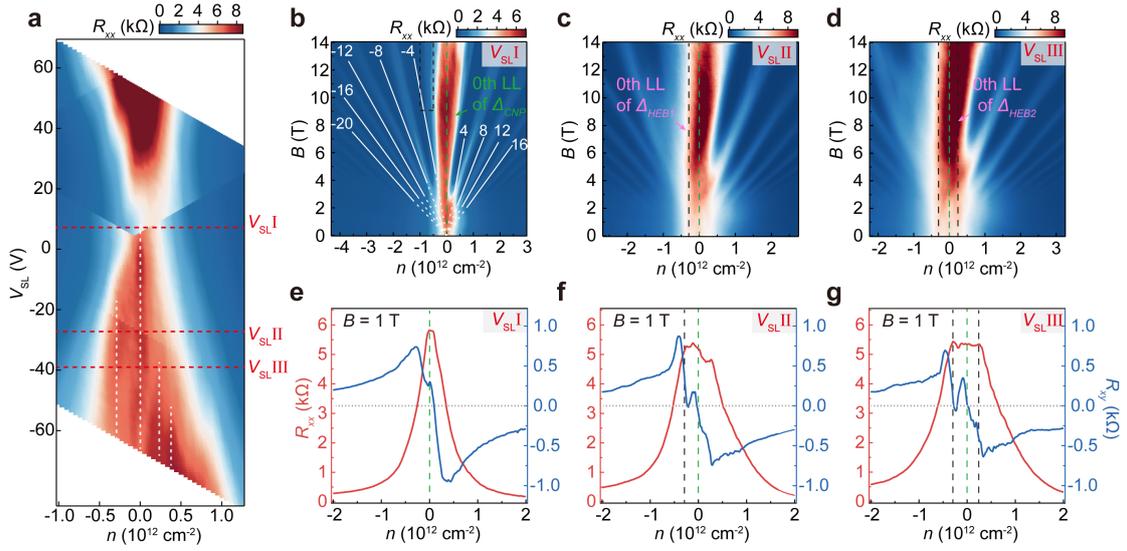

**Figure 2 | Transport characterization of the bilayer graphene Kagome superlattice at $T$ = 100 mK. a**, Longitudinal resistance ($R_{xx}$) map as a function of carrier density ($n$) and superlattice potential ($V_{SL}$). Multiple resistance maxima (white dashed lines) develop with increasing $V_{SL}$, reflecting tunable band-structure modulation. Note that the apparent color discontinuities near $V_{SL} \approx \pm 20$V are non-intrinsic artifacts arising from the construction of the map; see Supplementary Information for details. **b**, $R_{xx}$ map versus $n$ and magnetic field ($B$) at $V_{SL}$ = 7 V. A single Landau fan emanates from the charge-neutrality point (CNP), with the Landau level filling factors indicated by white lines. The zeroth Landau level (0th LL) is indicated by the green dashed line. The black dashed box highlights the region analyzed in Fig. 3.**c-d**, $R_{xx}$ map at $V_{SL}$ = -27 V (c) and -39 V (d), respectively, showing additional Landau fans originating from higher-energy band gaps ($\Delta_{HEB1}$ and $\Delta_{HEB2}$). Their 0th LLs are marked by black dashed lines. **e**, $R_{xx}$ and Hall resistance ($R_{xy}$) as a function of $n$ at $B$ = 1 T for $V_{SL}$ = 7 V. $R_{xx}$ shows a single resistance maximum and $R_{xy}$ shows sign change near the CNP. **f**, $R_{xx}$ and $R_{xy}$ at $B$ = 1 T for $V_{SL}$ = -27 V, exhibiting additional resistance maxima and modified $R_{xy}$ slopes. **g**, $R_{xx}$ and $R_{xy}$ at $B$ = 1 T for $V_{SL}$ = -39 V. Multiple $R_{xx}$ peaks and steep $R_{xy}$ slopes indicate complex band-structure evolution under strong Kagome-lattice modulation.

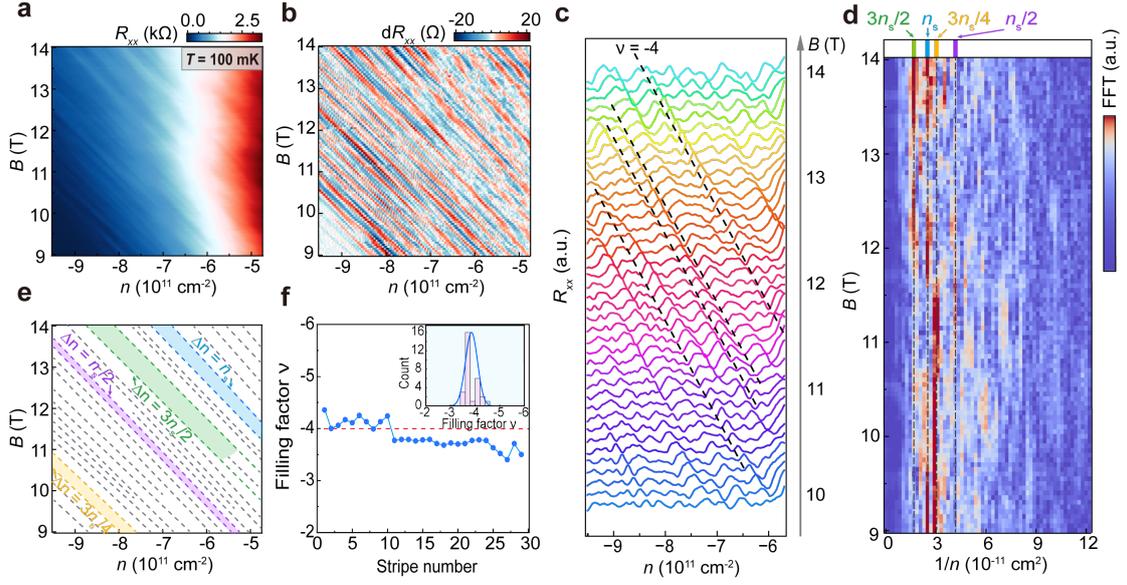

**Figure 3 | Signatures of a stack of correlated insulating states in the bilayer graphene Kagome superlattice at T=100 mK. a,** High-resolution $R_{xx}$ map extracted from Fig. 2b, focusing on a carrier-density range around $-7 \times 10^{11}$ cm$^{-2}$. **b,** Differential map of the background-subtracted $R_{xx}$ along the carrier-density axis, revealing multiple parallel minima. **c,** Linecuts of $R_{xx}$ at selected magnetic fields, showing linear trajectories of the minima with identical slopes (dashed guide lines). **d,** Fast Fourier transform (FFT) spectrum of panel (b), displaying four distinct carrier-density periods corresponding to $\frac{n_s}{2}$, $\frac{3n_s}{4}$, $n_s$, and $\frac{3n_s}{2}$. **e,** Schematic Landau level structure corresponding to the features in panel (b); a few representative intervals are highlighted to illustrate the four characteristic carrier-density periods. **f,** Landau-level filling factors extracted from the slopes in panel (e). The histogram inset yields an average filling factor of $\nu \approx -4$, confirming that the $R_{xx}$ minima originate from Landau levels at the same filling factor but different carrier densities.

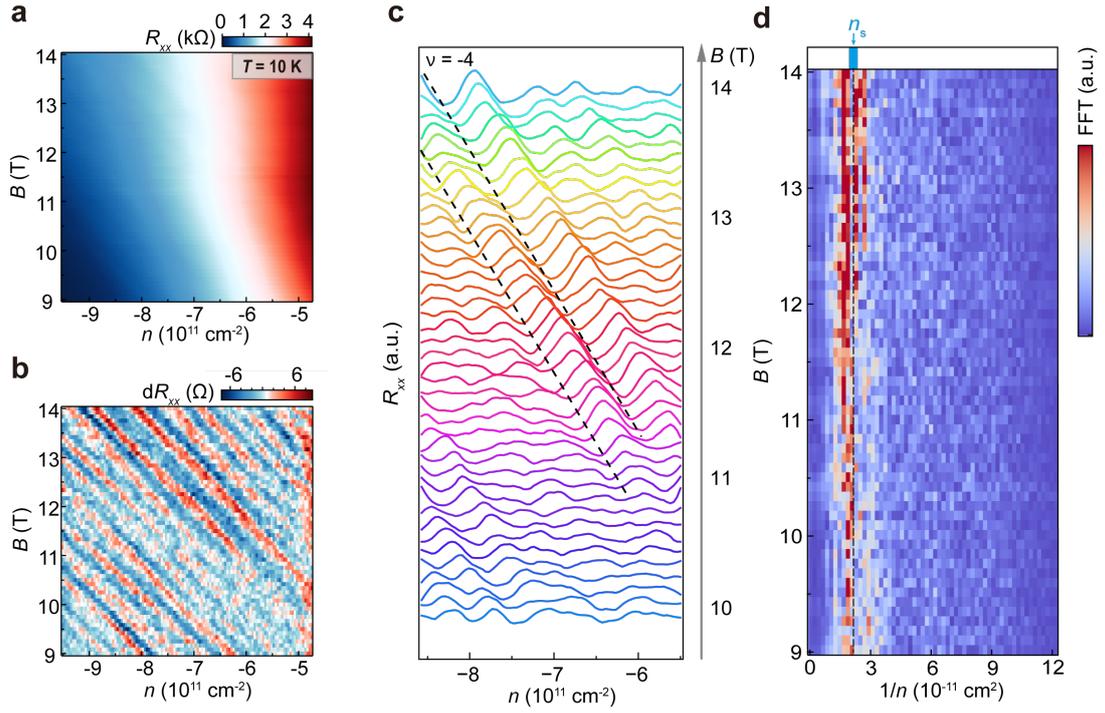

**Figure 4 | Suppression of correlated insulating states at elevated temperature (T = 10 K). a**, Longitudinal resistance ($R_{xx}$) map as a function of carrier density ($n$) and magnetic field ($B$) under the same $V_{SL}$ conditions as Fig. 3a. **b**, Differential map of the background-subtracted $R_{xx}$ along $n$, showing broader spacing between minima compared with the 100 mK data. **c**, $R_{xx}$ linecuts at representative magnetic fields; black dashed lines mark the aligned minima. **d**, Fast Fourier transform (FFT) of panel (b), revealing a single dominant carrier-density period of $(4.3 \pm 0.2) \times 10^{10}$ cm$^{-2}$, close to the superlattice filling density $n_s$, indicating thermal suppression of correlated insulating states.

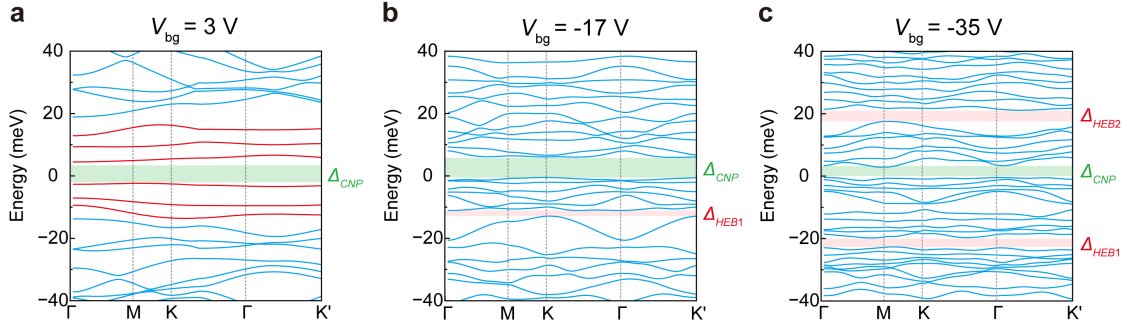

**Figure 5 | Calculated band structures of bilayer graphene under different Kagome superlattice potentials. a–c,** Continuum-model calculations of the reconstructed band structures of bilayer graphene (BLG) under the Kagome superlattice potential at various back-gate voltages ($V_{bg}$). (a) At $V_{bg}$ = 3 V, well-isolated flat bands (highlighted in red) emerge near the charge-neutrality point (CNP), with the corresponding energy gap ($\Delta_{CNP}$) indicated by the green shaded region. (b) At $V_{bg}$ = –17 V, the stronger modulation induces additional band reconstruction, with a higher-energy band gap ($\Delta_{HEB1}$, red shading) and the CNP gap ($\Delta_{CNP}$, green shading). (c) At $V_{bg}$ = –35 V, multiple higher-energy gaps ($\Delta_{HEB1}$, $\Delta_{HEB2}$) appear alongside the persistent CNP gap ($\Delta_{CNP}$), reflecting enhanced band overlap and complexity at large superlattice potentials.